\documentclass[journal,onecolumn]{IEEEtran}
\usepackage{amsmath}
\usepackage{booktabs}
\usepackage{latexsym}
\usepackage{mathrsfs}
\usepackage{amsthm}
\usepackage{amssymb}
\usepackage{amsfonts}
\usepackage{amsbsy}

\usepackage[shortlabels]{enumitem}
\usepackage{url}
\usepackage{array}
\usepackage{pdflscape}
\usepackage{xcolor}
\usepackage{stmaryrd}
\usepackage{verbatim}
\usepackage[noadjust]{cite}
\usepackage{graphicx}
\usepackage{tikz}

\theoremstyle{plain}
\newtheorem{thm}{Theorem}
\newtheorem{lem}[thm]{Lemma}

\newtheorem{defn}{Definition}
\newtheorem{example}{Example}
\newtheorem{remark}{Remark}

\usepackage[colorlinks=true]{hyperref}

\begin{document}
\title{A Construction of Asymptotically Optimal Cascaded CDC Schemes via Combinatorial Designs}

\author{\centerline{Yingjie Cheng, Gaojun Luo, Xiwang Cao, Martianus Frederic Ezerman, and San Ling}
\thanks{Y. Cheng, and X. Cao are with the Department of Mathematics, Nanjing University of Aeronautics and Astronautics, Nanjing 210016, China, and also with Key Laboratory of Mathematical Modeling and High Performance Computing of Air Vechicles (NUAA), MIIT, Nanjing 210016, China, e-mails: $\{\rm xwcao,chengyingjie\}$@nuaa.edu.cn}
\thanks{G. Luo, M. F. Ezerman, and S. Ling are with the School of Physical and Mathematical Sciences, Nanyang Technological University, 21 Nanyang Link, Singapore 637371, e-mails: $\{\rm gaojun.luo, fredezerman, lingsan\}$@ntu.edu.sg.}
\thanks{G. Luo, M. F. Ezerman, and S. Ling are supported by Nanyang Technological University Research Grant No. 04INS000047C230GRT01. X. Cao, Y. Cheng, and G. Luo are also supported by the National Natural Science Foundation of China under Grant 12171241.}
}

\maketitle

\begin{abstract}
A coded distributed computing (CDC) system aims to reduce the communication load in the {\tt MapReduce} framework. Such a system has $K$ nodes, $N$ input files, and $Q$ {\tt Reduce} functions. Each input file is mapped by $r$ nodes and each {\tt Reduce} function is computed by $s$ nodes. The objective is to achieve the maximum multicast gain. 
There are known CDC schemes that achieve optimal communication load. In some prominent known schemes, however, $N$ and $Q$ grow too fast in terms of $K$, greatly reducing their gains in practical scenarios. To mitigate the situation, some asymptotically optimal cascaded CDC schemes with $r=s$ have been proposed by using symmetric designs. In this paper, we put forward new asymptotically optimal cascaded CDC schemes with $r=s$ by using $1$-designs. Compared with earlier schemes from symmetric designs, ours have much smaller computation loads while keeping the other relevant parameters the same. We also obtain new asymptotically optimal cascaded CDC schemes with more flexible parameters compared with previously best-performing schemes.
\end{abstract}


\section{Introduction}
Distributed computing was envisioned as a tool for applications in data processing and analytics to process large amount of data efficiently. In \cite{dean2008}, Dean and Ghemawat introduced the popular distributed computing framework {\tt MapReduce}. As the label suggests, it decomposes into two stages, namely {\tt Map} and {\tt Reduce}. The need to exchange large amount of data among the computing nodes significantly reduces the system's performance. Li, Maddah-Ali, Yu, and Avestimehr in \cite{li2017} proposed \emph{coded distributed computing} (CDC) scheme to reduce the communication load by increasing the computation load of the {\tt Map} functions.

\subsection{Coded Distributed Computing Scheme}

A general $(K,N,Q,r,s)$ CDC scheme has $K$ computing nodes, $N$ input data files of equal size, and $Q$ output values. Each of the values is computed based on the $N$ input data files. It is typical that $N$ is a large positive integer $\geq K$. In such a scheme, the computation is divided into three phases, namely the {\tt Map}, {\tt Shuffle} and {\tt Reduce} phases. In the {\tt Map} phase, any distinct $r$-subset of the $K$ computing nodes exclusively maps each input file to $Q$ intermediate values (IVs) of size $T$ bits each. During the {\tt Shuffle} phase, each {\tt reduce} function is assigned to an $s$-subset of the computing nodes. Each computing node can then generate coded symbols from its local IVs to all other computing nodes. In the {\tt Reduce} phase, each computing node runs each {\tt Reduce} function that has been assigned to the node, after receiving the coded signals during the {\tt Shuffle} phase. As identified by Chowdhury {\it et al.} in \cite{chowdhury2011}, the {\tt Shuffle} phase creates a substantial communication bottleneck. In this phase, nodes spend most of their execution time in exchanging IVs among themselves. To reduce the execution time in the Shuffle phase, one can benefit from a fundamental tradeoff between the \emph{computation load} in the {\tt Map} phase and the \emph{communication load} in the {\tt Shuffle} phase. Li {\it et al.} had already formulated and characterized the tradeoff in \cite{li2017}. 

If $s=1$, then a $(K,N,Q,r,s)$ CDC scheme can be seen as a coded caching scheme for the D2D network discussed, {\it e.g.}, in \cite{li2017} and \cite{ji2015}. Whenever $s>1$, a CDC scheme is \emph{cascaded}, where each {\tt Reduce} function is calculated by multiple nodes. Zhao {\it et al.} in \cite{zhao2018} and, subsequently, Chen and Sung in \cite{chen2022} have looked into security issues in data shuffling and constructed \emph{weakly} secure CDC schemes. Studies on CDC have expanded in various directions. Prominent works on distributed computing in heterogeneous networks include those done by Kiamari, Wang, and Avestimehr in \cite{kiamari2017}, by Shakya, Li, and Chen in \cite{shakya2018}, and by Woolsey, Chen, and Ji in \cite{woolsey2021combinatorial}, \cite{woolsey2019}, \cite{woolsey2020coded}, and \cite{xu2019}. In wireless network, one finds the work of Li, Chen, and Wang in \cite{li2019} and that of Li, Yu, Maddah-Ali, and Avestimehr in \cite{li2016edge}. Lee, Suh, and Ramchandran in \cite{lee2017high} and D'Oliviera, El Rouayheb, Heinlein, and Karpuk in \cite{d2020notes} investigated coded matrix multiplication. 

\subsection{Known Results on Cascaded Coded Distributed Computing Schemes}

The authors of \cite{li2017} constructed a cascaded CDC scheme with $N=\binom{K}{r}$ input files and $Q=\binom{K}{s}$ output functions. The average number of nodes that store each file is $r$ and the average number of nodes that calculate each function is $s$. They proved that their scheme is \emph{optimal} as its communication load reaches the theoretical minimum. The parameters $N$ and $Q$ in their scheme, however, grow too fast in $K$. This greatly reduces the scheme's practical gains, as explained by Konstantinidis and Ramamoorthy in \cite{kon2020}. 

Woolsey, Chen, and Ji in \cite{woolsey2021} proposed a combinatorial structure, called the \emph{hypercube structure}, to design the file and functions assignments. This scheme splits the data set into $N=\left(\frac{K}{r}\right)^{r-1}$ files and defines $Q=\left(\frac{K}{r}\right)^{r-1}$ output functions. They established that their scheme's communication load is close to that of the scheme in \cite{li2017}. Unfortunately, as $r$ increases, the $N$ and $Q$ in their scheme start to grow too fast in $K$. 

In \cite{jiang2020}, Jiang and Qu presented a cascaded CDC scheme with $N=\left(\frac{K}{r}\right)^{r-1}$ and $Q=\frac{K}{\gcd(K,s)}$ by using \emph{placement delivery array}, which had been previously introduced to construct coded caching schemes by Yang, Cheng, Tang, and Chen in \cite{yan2017}. The scheme, however, is \emph{not} asymptotically optimal. Its communication load is approximately twice as large as that of the scheme in \cite{li2017}. Very recently, Jiang, Wang, and Zhou used symmetric balanced incomplete block design (SBIBD) to generate the data placement and the {\tt Reduce} function assignments to come up with an asymptotically optimal scheme with $K=N=Q$ in \cite{jiang2022}. The combinatorial route was also used by Cheng, Wu, and Li in \cite{cheng2023} to devise some asymptotically optimal schemes based on $t$-designs, with $t\geq 2$, and $t$-GDDs (group divisible designs). In \cite{cheng20231}, Cheng, Luo, Cao, Ezerman, and Ling constructed two asymptotically optimal cascaded CDC schemes with $K=N=Q$ via combinatorial designs. The first one uses symmetric designs for the case $r+s=K$ to achieve a lower communication load than in the scheme of \cite{jiang2022}. The second construction uses $1$-designs from almost difference (AD) sets.

Table \ref{table1} lists the above-mentioned known cascaded CDC schemes and their respective parameters. 
\begin{table}
\caption{Cascaded CDC Schemes: Known and New}
\label{table1}
\renewcommand{\arraystretch}{1.2}
\centering
\begin{tabular}{c|l|c|c|c|c|c|l}
\toprule
Ref. & Parameters are & Number of & Computation & Replication & Number of & Number of Reduce & Communication \\
& Positive Integers & Nodes $K$ & Load $r$ & Factor $s$ & Files $N$ & Functions $Q$ & Load $L$ \\
\toprule

\cite{li2017} &  $K,r,s$, with & $K$ & $r$ & $s$ & $\binom{K}{r}$ & $\binom{K}{s}$ & $\sum^{\min\{r+s,K\}}_{\ell=\max\{r+1,s\}}$ \\
& $1\leq r,s\leq K$ &&&&&& $\frac{\binom{K-r}{K-\ell}\binom{r}{\ell-s}}{\binom{K}{s}}\frac{\ell-r}{\ell-1}$\\

\midrule
\cite{woolsey2021} & $K,r$, with & $K$ & $r$ & $r$ & $\left(\frac{K}{r}\right)^{r-1}$ & 
$\left(\frac{K}{r}\right)^{r-1}$ & $\frac{r^{r}(k-r)}{K^{r}(r-1)} + \sum^{r}_{\ell=2}$ \\
& $1\leq r\leq K$&&&&&& $\left(\frac{r}{K}\right)^{r+\ell} \, \binom{r}{\ell} 
\, \binom{K/2}{2}^{\ell} \frac{2^{\ell} \ell}{2 \ell-1}$\\

\midrule

\cite{jiang2020} & $K,r,s$, with & $K$ & $r$ & $s$ & $\left(\frac{K}{r}\right)^{r-1}$ & $\frac{K}{\gcd(K,s)}$ & $\frac{s}{r-1} \, \left(1-\frac{r}{K}\right)$ \\
& $1\leq r,s\leq K$&&&&&& \\

\midrule

\cite{jiang2022} & $(K,r,\lambda)$~SD & $K$ & $r$ & $r$ & $K$ & $K$ & $\frac{r \, (K-r)}{(r-1) \, K}$ \\
\cline{3-8}
& with $2\leq \lambda \leq r$ &$K$&$r$&$K-r$&$K$&$K$&$\frac{K-r}{K-1}$ \\

\midrule

\cite{cheng2023} & $(N,M,\lambda)$ $t$-design & $\frac{\lambda \binom{N}{t}}{\binom{M}{t}}$ &$\frac{\lambda \binom{N-1}{t-1}}{\binom{M-1}{t-1}}$ & $\frac{\lambda \binom{N-1}{t-1}}{\binom{M-1}{t-1}}$ & $N$ & $N$ & $\frac{N-1}{2\,N}$ \\
& with $t\geq 2$ &&&&&& \\

\midrule

\cite{cheng2023} & $(m,q,M,\lambda)$ $t$-GDD & $\frac{\lambda\binom{m}{t} \, q^{t}}{\binom{M}{t}}$ &$\frac{\lambda \binom{m-1}{t-1} \, q^{t-1}}{\binom{M-1}{t-1}}$ & $\frac{\lambda \binom{m-1}{t-1} \, q^{t-1}}{\binom{M-1}{t-1}}$
& $m \, q$ & $m \, q$ & $\frac{1}{2}+\frac{q-2}{m \, q}$ \\
\midrule
\cite{cheng20231} & 
$(K,r,\lambda)$ SD &$K$&$r$&$K-r$&$K$&$K$&$\frac{(K-1)^{2}-rK+K}{K(K-1)}$ \\ 
& with $1\leq \lambda \leq r$&  
&&&&& \\
\midrule
\cite{cheng20231} & $(n,k,\lambda,\mu)$ AD set & $n$ & $k$ & $k$ & $n$ & $n$ & $\frac{n-1}{2n}$ \\
& with $1\leq\lambda\leq k-1$&&&&& \\
\midrule
\cite{cheng20231} & $(n,k,0,\mu)$ AD set & $n$ & $k$ & $k$ & $n$ & $n$ & $\frac{2n-2-k(k-1)}{2n}$ \\
\bottomrule
\toprule
New & $K,r \mbox{ with } 3r \geq 2n$ & $K$ & $r$ & $r$ & $K$ & $K$ & $\frac{K-r}{K}$ \\
\bottomrule
\end{tabular}
\end{table}

\subsection{Our Contributions}

Our focus here is on the cascaded CDC schemes with $r=s$. We carefully design the data placement and the {\tt Reduce} function assignment using specific $1$-designs. Based on well-chosen $1$-designs, we construct a new class of asymptotically optimal cascaded CDC schemes. When compared with previously known schemes in Table \ref{table1}, ours has the following advantages.
\begin{itemize}
\item Compared with the Li-CDC scheme in \cite{li2017}, ours has \emph{much smaller $N$ and $Q$}, that is, ours requires much smaller number of input files and {\tt Reduce} functions. When the number of nodes $K$ is large enough, the communication loads of our scheme approximates that of the scheme in \cite{li2017}.
\item Compared with the Jiang-CDC scheme in \cite{jiang2022}, ours has a \emph{smaller communication load} for the same parameters $(K,N,Q,r,s)$. Using SBIBD, which we will soon recall in Section \ref{sec:prelims} as a special type of $2$-design, we construct \emph{new CDC schemes with more flexible parameters} than those in the scheme in \cite{jiang2022}.
\item Compared with the scheme in \cite{cheng2023,cheng20231}, ours has a \emph{smaller communication load} for the same parameters $(K,N,Q,r,s)$.
\end{itemize}

In terms of organization, Section \ref{sec:prelims} recalls useful properties of combinatorial designs and cascaded coded distributed computing systems. We provide the main results and illustrate our schemes with examples in Section \ref{sec:main}. Comparative performance of our schemes and related prior schemes can be found in Section \ref{sec:perform}. Concluding remarks in Section \ref{sec:conclude} wraps the main part of the paper up. Appendix A provides a detailed proof of Theorem \ref{thm1} whereas Appendix B supplies our proof of Lemma \ref{lem3.1}.

\section{Preliminaries}\label{sec:prelims}
We denote by $\vert \cdot \vert$ the cardinality of a set or the length of a vector. For any integers $a$ and $b$ with $a<b$, we use $[a,b]$ to denote the set $\{a,a+1,\ldots,b\}$. If $a=1$, then we use the shortened form $[b]$.

\subsection{Cascaded Coded Distributed Computing System}
\begin{figure}[b!]
\center
\caption{A two-stage distributed computing framework. The overall computation is decomposed into computing a set of {\tt Map} and {\tt Reduce} functions}
\label{fig1}
\includegraphics[width=0.7\linewidth]{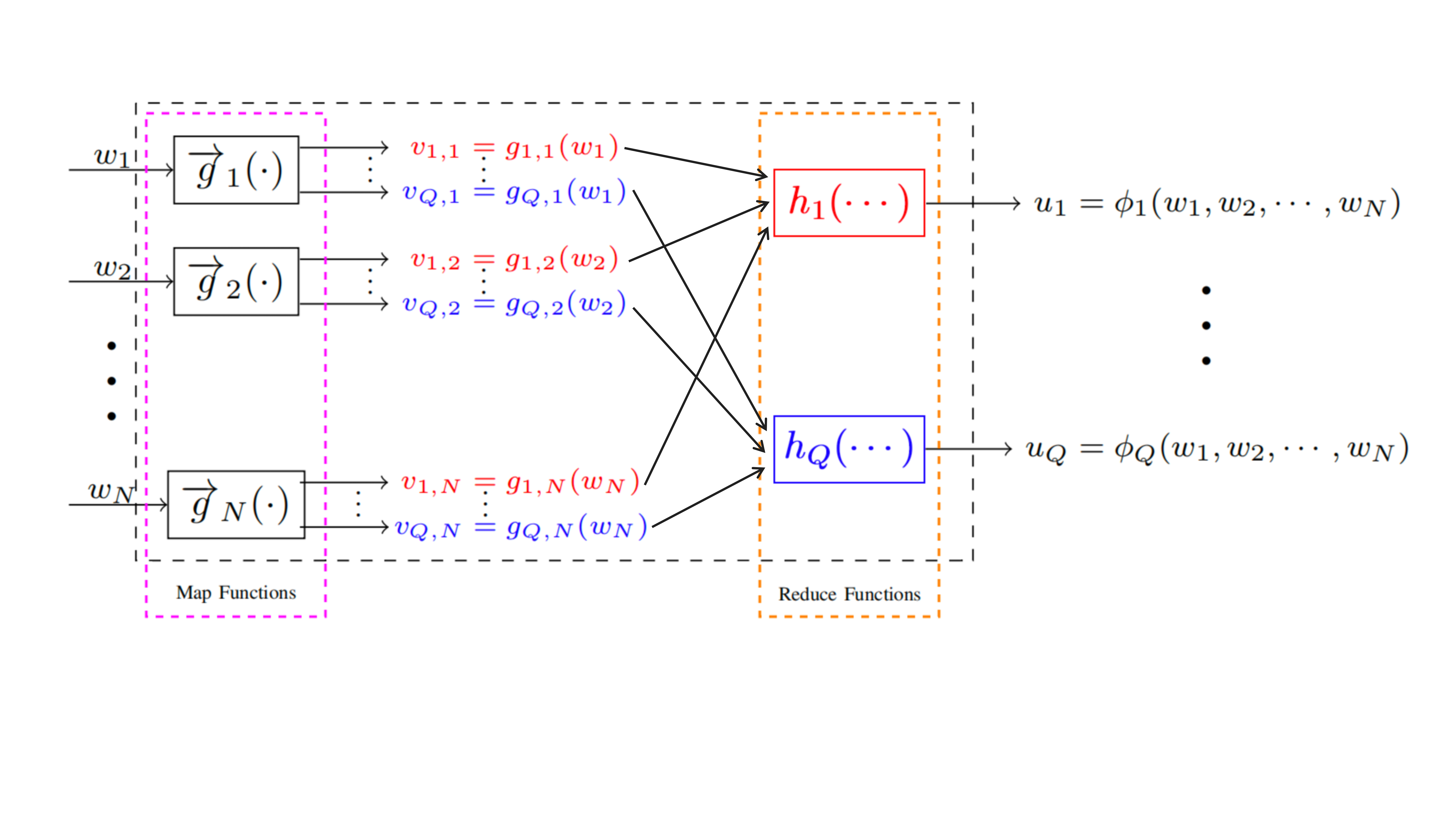}
\end{figure}

In a coded distributed computing system, there are $K$ distributed computing nodes to compute $Q$ {\tt Reduce} functions by taking advantage of $N$ input files, each of equal size, where $N \geq K$. Let $W=\{w_{1},w_{2},\ldots,w_{N}\}$ be the set of the $B$-bit $N$ files. The set of $Q$ functions is $\mathcal{Q} = \{\phi_{1},\phi_{2},\ldots,\phi_{Q}\}$, where, for any $q\in[Q]$, $\phi_{q}$ maps the $N$ files to a $C$-bit value $u_{q}=\phi_{q}(w_{1},w_{2},\ldots,w_{N})\in \mathbb{F}_{2^{C}}$. Figure \ref{fig1} shows how each output function $\phi_{q} \, : \, q \in[Q]$ decomposes as 
\[
\phi_{q}(w_{1},w_{2},\ldots,w_{N}) = h_{q} \left(g_{q,1}(w_{1}),g_{q,2}(w_{2}),\ldots,g_{q,N}(w_{N})\right),
\]
where $g_{q,n}$ is a {\tt Map} function, for any $q\in[Q]$ and $n\in[N]$, and $h_{q}$ is a {\tt Reduce} function for any $q\in[Q]$. The intermediate value (IV) is $v_{q,n} := g_{q,n}(w_{n})$, where $q\in[Q]$ and $n\in[N]$. A cascaded CDC scheme consists of the following three phases.

\begin{itemize}
\item {\tt Map} {\bf Phase}.\\
Each node $k\in \mathcal{K}$ first stores $M$ files, which we collect into the set $\mathcal{Z}_{k}$. For each file $w_{n}$, let $\mathcal{D}_{n}$ represent the node set each of which stores file $w_{n}$. The files stored by node $k$ are elements of the set 
\begin{equation} \label{eq0.1}
\mathcal{Z}_{k}=\{w_{n} \, : \, n\in[N] \mbox{ and } k\in \mathcal{D}_{n}\}.
\end{equation}
Using the stored files in (\ref{eq0.1}) and the {\tt Map} functions $\{g_{q,n}(\cdot)\}$, node $k$ is exclusively mapped to the IVs in the set
\[
\mathcal{I}_{k}=\{v_{q,n} = g_{q,n}(w_{n}) \, : \, q\in[Q],n\in[N] \mbox{, and } k\in \mathcal{D}_{n}\}.
\]
\item  {\tt Shuffle} {\bf Phase}.\\
Each node $k\in \mathcal{K}$ is associated with the set of output functions
\begin{equation} \label{eq0.2}
\mathcal{Q}_{k}=\{\phi_{q} \, : \, q\in[Q] \mbox{ and } k\in \mathcal{A}_{q}\},
\end{equation}
where $\mathcal{A}_{q}$ represents the node set, each of which is assigned to compute $\phi_{q}$. Consequently, each node needs to exchange its calculated IVs with all other nodes. Each node $k \, : \, k\in \mathcal{K}$ multicasts a coded message $X_{k}$ of length $\ell_{k}$. At the end of this phase, we assume that each node would have received the messages from all other nodes error-free. 
\item {\tt Reduce} {\bf Phase}.\\
Upon receiving the coded signals in the set $\mathcal{X}=\{X_{1},X_{2},\ldots,X_{K}\}$, each node $k$ can compute each {\tt Reduce} function in $\mathcal{Q}_{k}$ by using its locally computed IVs in $\mathcal{I}_{k}$. 
\end{itemize}

Following \cite{li2017}, we adopt two important measures on CDC schemes. The first one is the \emph{computation load} 
\[
r=\frac{\sum^{K}_{k=1} \vert\mathcal{W}_{k}\vert}{N},
\]
which measures the average number of nodes that store each file. The second one is the \emph{communication load} 
\[
L=\frac{\sum^{K}_{k=1} \ell_{k}}{Q \cdot N \cdot T},
\] 
which measures the ratio of the amount of transmitted data to the quantity $Q \cdot N \cdot T$. The objective is of course to design schemes with the least value of $L$ for a given $r$. The optimal communication load has already been established and is reproduced here as the next lemma for convenience.
\begin{lem} \label{lem0.1} \cite{li2017}
Let $r$ be the given computation load and let $s$ be the number of nodes that calculate each function. If $K$ is a positive integer, then, for $r,s\in[K]$, there exists a CDC scheme that achieves the optimal communication load
\[
L = \sum^{\min\{r+s,K\}}_{\ell=\max\{r+1,s\}} \frac{\binom{K-r}{K-\ell}\binom{r}{\ell-s}}{\binom{K}{s}}\frac{\ell-r}{\ell-1}.
\]
\end{lem}
Since the CDC scheme in Lemma \ref{lem0.1} is the Li-CDC scheme, we denote its communication load by $L_{\rm Li}$.

\subsection{Structures from Combinatorial Designs}
The next two definitions, of combinatorial design and $t$-design, respectively, are reproduced from \cite{col2007}. 
\begin{defn} 
A design is a pair $(\mathcal{X},\mathfrak{B})$ with the following properties.
\begin{itemize}
\item The elements of the set $\mathcal{X}$ are called points. 
\item $\mathfrak{B}$ is a collection of nonempty subsets of $\mathcal{X}$ called blocks.
\end{itemize}
\end{defn}

\begin{defn}\label{Def1} Let $n$, $K$, $k$, $t$, and $\lambda$ be positive integers. A $t$-$(n,k,\lambda)$ design is an $(\mathcal{X},\mathfrak{B})$ design whose $\mathcal{X}$ contains $n$ points and $\mathfrak{B}$ has $K$ blocks that meet the following properties.
\begin{itemize}
\item $\vert\mathcal{B}\vert = k$ for any $\mathcal{B}\in\mathfrak{B}$.
\item Every $t$-subset of $\mathcal{X}$ is contained in exactly $\lambda$ blocks.
\end{itemize}
\end{defn}

Based on Definition \ref{Def1}, the number of blocks is
\[
K=\frac{\lambda{n}{t}}{{k}{t}}.
\]
It is immediate to see that a $t$-$(n,k,\lambda)$ design is also a $t'$-$(n,k,\lambda')$ design for any $t'\leq t$ and 
\[
\lambda'= \frac{\lambda \binom{n-t'}{t-t'}}{\binom{k-t'}{t-t'}}.
\]

\section{Main Result}\label{sec:main}
We now propose a new cascaded CDC scheme for the case of $r=s$. Our proof of Theorem \ref{thm1} is rather lengthy. To maintain our focus on the constructed scheme, the proof is moved to the appendix.
\begin{thm} \label{thm1}
Given positive integers $n$ and $t$ with $3t \geq 2n$, one can construct a CDC scheme, with $n$ distributed computing nodes, $N=n$ input files, and $Q=n$ output functions, that satisfies the following conditions.
\begin{itemize}
  \item Each output function is computed by $s=t$ nodes.
  \item The computation load is $r=t$.
  \item The communication load is $L=\frac{n-t}{n}$.
\end{itemize}
\end{thm}

We use $n=6$ and $t=4$ in the next example to illustrate our construction.
\begin{example}
When $N=Q=K=6$, we have $\mathcal{W}=\{w_{0},w_{1},\ldots,w_{5}\}$ and the set of output functions $\mathcal{Q}=\{\phi_{0},\phi_{1},\ldots,\phi_{5}\}$. In the {\tt Map} phase, the nodes store the respective corresponding files
\begin{align*}
\mathcal{Z}_{\mathcal{B}_{0}}&=\{w_{0},w_{1},w_{2},w_{3}\}, &\mathcal{Z}_{\mathcal{B}_{1}} &=\{w_{1},w_{2},w_{3},w_{4}\}, &\mathcal{Z}_{\mathcal{B}_{2}}&=\{w_{2},w_{3},w_{4},w_{5}\},\\
\mathcal{Z}_{\mathcal{B}_{3}} &=\{w_{3},w_{4},w_{5},w_{0}\}, &\mathcal{Z}_{\mathcal{B}_{4}} &=\{w_{4},w_{5},w_{0},w_{1}\}, &\mathcal{Z}_{\mathcal{B}_{5}}&=\{w_{5},w_{0},w_{1},w_{2}\}.
\end{align*}
The computation load is, therefore, $r=\frac{4\cdot 6}{6}=4$.

Let the assignment of the {\tt Reduce} functions relative to the workers be
\begin{align*}
\mathcal{Q}_{\mathcal{B}_{0}} &=\{\phi_{0},\phi_{1},\phi_{2},\phi_{3}\}, &\mathcal{Q}_{\mathcal{B}_{1}} &=\{\phi_{1},\phi_{2},\phi_{3},\phi_{4}\}, &\mathcal{Q}_{\mathcal{B}_{2}}&=\{\phi_{2},\phi_{3},\phi_{4},\phi_{5}\},\\
\mathcal{Q}_{\mathcal{B}_{3}}&= \{\phi_{3},\phi_{4},\phi_{5},\phi_{0}\}, & \mathcal{Q}_{\mathcal{B}_{4}} &=\{\phi_{4},\phi_{5},\phi_{0},\phi_{1}\}, &\mathcal{Q}_{\mathcal{B}_{5}}&=\{\phi_{5},\phi_{0},\phi_{1},\phi_{2}\}.
\end{align*}
Each function is computed by $s=4$ nodes.

The nodes can locally compute the respective IVs 
\begin{align*}
\mathcal{I}_{\mathcal{B}_{0}} &=\{v_{q,n}\mid q\in\{0,1,2,3,4,5\}, n\in\{0,1,2,3\}\}, &\mathcal{I}_{\mathcal{B}_{1}}&=\{v_{q,n}\mid q\in\{0,1,2,3,4,5\}, n\in\{1,2,3,4\}\},\\
\mathcal{I}_{\mathcal{B}_{2}}&=\{v_{q,n}\mid q\in\{0,1,2,3,4,5\}, n\in\{2,3,4,5\}\}, &\mathcal{I}_{\mathcal{B}_{3}} &=\{v_{q,n}\mid q\in\{0,1,2,3,4,5\}, n\in\{3,4,5,0\}\},\\
\mathcal{I}_{\mathcal{B}_{4}}&=\{v_{q,n}\mid q\in\{0,1,2,3,4,5\}, n\in\{4,5,0,1\}\}, 
&\mathcal{I}_{\mathcal{B}_{5}}&=\{v_{q,n}\mid q\in\{0,1,2,3,4,5\}, n\in\{5,0,1,2\}\}.
\end{align*}
Table \ref{table2.1} lists the IVs required by the nodes.
\begin{table}[h!]
\caption{Intermediate values $\{v_{q,n}\}$ required by workers in $\mathfrak{B}$}
\label{table2.1}
\centering
\begin{tabular}{c|cccccc}
\toprule
Parameters & \multicolumn{6}{c}{worker set $\mathfrak{B}$} \\
\midrule
 & $\mathcal{B}_{0}$ & $\mathcal{B}_{1}$ & $\mathcal{B}_{2}$ & $\mathcal{B}_{3}$ & $\mathcal{B}_{4}$ & $\mathcal{B}_{5}$ \\
\midrule
$q$ & $0,1,2,3$ & $1,2,3,4$ & $2,3,4,5$ & $3,4,5,0$ & $4,5,0,1$ & $5,0,1,2$  \\
\midrule
$n$ & $4,5$ & $5,0$ & $0,1$ & $1,2$ & $2,3$ & $3,4$ \\
\bottomrule
\end{tabular}
\end{table}

Taken collectively, the nodes can send the coded signals according to Table \ref{table2.2}, where $a_{1}\in \mathbb{F}_{2^{T}} \setminus \{1\}$. 

\begin{table}[h!]
\caption{Coded signals sent by workers in $\mathfrak{B}$}
\label{table2.2}
\centering
\begin{tabular}{ccccccc}
\toprule
$\mathcal{B}_{0}$ & $\mathcal{B}_{1}$ & $\mathcal{B}_{2}$ \\
\midrule
$v_{4,0}+a_{1}v_{4,1}+v_{4,2}+v_{4,3}$ & $v_{5,1}+a_{1}v_{5,2}+v_{5,3}+v_{5,4}$ & $v_{0,2}+a_{1}v_{0,3}+v_{0,4}+v_{0,5}$ \\
$v_{5,0}+v_{5,1}+v_{5,2}+v_{5,3}$ & $v_{0,1}+v_{0,2}+v_{0,3}+v_{0,4}$ & $v_{1,2}+v_{1,3}+v_{1,4}+v_{1,5}$ \\
\midrule
\midrule
$\mathcal{B}_{3}$ & $\mathcal{B}_{4}$ & $\mathcal{B}_{5}$ \\
\midrule
$v_{1,3}+a_{1}v_{1,4}+v_{1,5}+v_{1,0}$ & $v_{2,4}+a_{1}v_{2,5}+v_{2,0}+v_{2,1}$ & $v_{3,5}+a_{1}v_{3,0}+v_{3,1}+v_{3,2}$ \\
$v_{2,3}+v_{2,4}+v_{2,5}+v_{2,0}$ & $v_{3,4}+v_{3,5}+v_{3,0}+v_{3,1}$  & $v_{4,5}+v_{4,0}+v_{4,1}+v_{4,2}$ \\
\bottomrule
\end{tabular}
\end{table}

Nodes $\mathcal{B}_{0}$ and $\mathcal{B}_{1}$, for instance, send the respective coded signals 
\[
x_{5,0} = v_{5,0}+v_{5,1}+v_{5,2}+v_{5,3} \mbox{ and } x_{5,1}=v_{5,1}+a_{1}v_{5,2}+v_{5,3}+v_{5,4}.
\]
Upon receiving $x_{5,0}$ and $x_{5,1}$, node $\mathcal{B}_{2}$ can decode the required IVs $v_{5,0}$ and $v_{5,1}$ by solving for
\[
\begin{cases}
v_{5,0}+v_{5,1} &=x_{5,0}-v_{5,2}-v_{5,3}; \\
v_{5,1} &=x_{5,1}-a_{1}v_{5,2}-v_{5,3}-v_{5,4}.
\end{cases}
\]
After getting $x_{5,0}$ and $x_{5,1}$, node $\mathcal{B}_{3}$ can decode the IVs $v_{5,1}$ and $v_{5,2}$ by solving for
\[
\begin{cases}
v_{5,1}+v_{5,2} &=x_{5,0}-v_{5,0}-v_{5,3}; \\
v_{5,1}+a_{1}v_{5,2} &=x_{5,1}-v_{5,3}-v_{5,4}.
\end{cases}
\]
On having $x_{5,0}$ and $x_{5,1}$, node $\mathcal{B}_{4}$ can decode the IVs $v_{5,2}$ and $v_{5,3}$ by finding the solution for
\[
\begin{cases}
v_{5,2}+v_{5,3} &=x_{5,0}-v_{5,0}-v_{5,1}; \\
a_{1}v_{5,2}+v_{5,3} &=x_{5,1}-v_{5,1}-v_{5,4}.
\end{cases}
\]
Having received $x_{5,0}$ and $x_{5,1}$, node $\mathcal{B}_{5}$ can decode the IVs $v_{5,3}$ and $v_{5,4}$ by solving for
\[
\begin{cases}
v_{5,3} &=x_{5,0}-v_{5,0}-v_{5,1}-v_{5,2}; \\
v_{5,3}+v_{5,4} &=x_{5,1}-v_{5,1}-a_{1}v_{5,2}.
\end{cases}
\]
Similarly, all other nodes can obtain their respective required IVs $v_{x,y}$ with $y\in\{0,1,2,3,4,5\}$ and $x\in\{0,1,2,3,4\}$. The communication load is $L=\frac{6 \cdot (6-4)}{6 \cdot 6}=\frac{1}{3}$.
\end{example}

\section{Comparative Performance}\label{sec:perform}
The Li-CDC scheme becomes less practical as $N=\binom{K}{r}$ and $Q=\binom{K}{s}$ grow fast in $K$. The steep increase in the number of input files deteriorates the scheme's performance. In Theorem \ref{thm1}, the numbers of input files and output functions in our new scheme are equal to the number of nodes. In this regard, our scheme is superior to the Li-CDC scheme. 

Let us recall the asymptotically optimal cascaded CDC scheme, with $r=s$, based on symmetric designs from \cite{jiang2022}. Due to the underlying combinatorial structures, for a prime power $b$, the scheme has 
\[
N=b^{2}+b+1, \quad Q=b^{2}+b+1, \quad r=s=b^{2}, \quad L=\frac{b^{2}}{(b^{2}+b+1)(b-1)}.
\]
Our new scheme has two advantages over that scheme. Ours has a smaller communication load for the same parameters $(K, N, r, Q)$ and has more flexible parameters. 

\begin{figure}[ht!]
\center
\caption{Comparison of the communication loads, in prime power $b$, between our scheme and the scheme in \cite{jiang2022}.}
\label{fig2}
\includegraphics[width=0.7\linewidth]{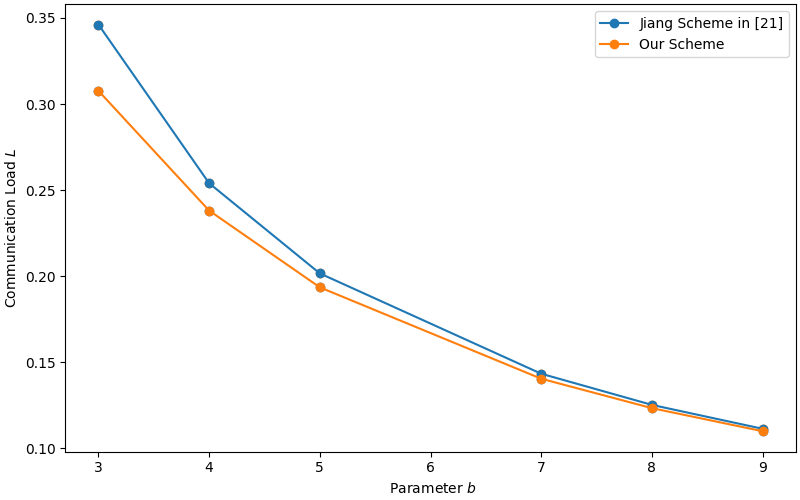}
\end{figure}

Constructed using symmetric designs with parameters $\left(b^{2}+b+1,b^{2},b^{2}-b\right)$, for a given prime power $b$, the scheme in \cite{jiang2022} has 
\[
r=s=b^{2} \mbox{ and } 
L_{\rm Jiang}=\frac{b^{2}(b+1)}{(b^{2}+b+1)(b^{2}-1)}.
\]
In our new scheme construction, letting $n=b^{2}+b+1$ and $t=b^{2}$, we get $3t\geq 2n$, whenever $b\geq 3$. By Theorem \ref{thm1}, the resulting CDC scheme has 
\[
r=s=b^{2} \mbox{ and } 
L_{\rm Our}=\frac{b+1}{b^{2}+b+1}.
\]
Since $\frac{b^{2}}{b^{2}-1}>1$, we confirm that $L_{\rm Our}<L_{\rm Jiang}$. The scheme based on the $(b^{2}+b+1,b^{2},b^{2}-b)$ SD in \cite{jiang2022} is known to be asymptotically optimal. Hence, letting $n=b^{2}+b+1$ and $t=b^{2}$, our cascaded CDC scheme is also asymptotically optimal. Figure \ref{fig2} gives a clear comparison between the two schemes.

Let $w,v_{1},v_{2},\ldots,v_{c}$ be $c+1$ distinct positive integers with $\lfloor\frac{w}{2}\rfloor\geq \max\{v_{1},v_{2},\ldots,v_{c}\}$. 
For any integer $p\geq 2$, by Theorem \ref{thm1}, we can construct a class of CDC schemes with $r=s=p^{w}$ and $K=p^{w}+ y$, where $y=\sum^{c}_{i=1}p^{v_{i}}$. It is immediate to confirm that the communication load is $L_{1}=\frac{y}{p^{w}+y}$ and that the numbers of input files and output functions are both $p^{w}+y$, which is also the value for $K$. We now show that our scheme is asymptotically optimal by establishing that $\frac{L_{1}}{L_{li}}$ converges to $1$ as $p$ goes to infinity. We use the following lemma, whose proof is in Appendix B.
\begin{lem} \label{lem3.1}
Let $p$ be a positive integer. If $p \to +\infty$, then 
\begin{equation}\label{for31}
\sum^{y}_{i=0} i \binom{y}{y-i} \binom{p^{w}}{i} > (y-2) \binom{p^{w}+y}{y}.
\end{equation}
\end{lem}

If $p\geq 3$, then $3p^{w}>2(p^{w}+y)$. Taking $r=s=p^{w}$ and $K=p^{w}+y$ in Lemma \ref{lem0.1} yields
\[
L_{\rm Li} = \sum^{p^{w}+y}_{\ell = p^{w}+1} \frac{\binom{y}{p^{w}+y-\ell} \binom{p^{w}}{\ell-p^{w}}} {\binom{p^{w}+y}{p^{w}}} 
\frac{\ell-p^{w}}{\ell-1} = 
\sum^{y}_{i=1}\frac{\binom{y}{i} \binom{p^{w}}{i}}{\binom{p^{w}+y}{p^{w}}} \frac{i}{p^{w}+i-1}.
\]
By Lemma \ref{lem3.1}, as $p\to +\infty$, we get
\begin{align*}
L_{\rm Li} &=\frac{1}{\binom{p^{w}+y}{p^{w}}}\sum^{y}_{i=1} \frac{i}{p^{w}+i-1} \binom{y}{i} \binom{p^{w}}{i} \\
&\geq \frac{1}{\binom{p^{w}+y}{p^{w}} 
(p^{w}+y-1)} \sum^{y}_{i=1} i \binom{y}{i} \binom{p^{w}}{i} \\
& >\frac{(y-2) \binom{p^{w}+y}{p^{w}}}{\binom{p^{w}+y}{p^{w}} (p^{w}+y-1)} = 
\frac{y-2}{p^{w}+y-1}.
\end{align*}
On the other hand,
\begin{align*}
L_{\rm Li}&= \frac{1}{\binom{p^{w}+y}{p^{w}}}\sum^{y}_{i=1} \frac{i}{p^{w}+i-1} \binom{y}{i} \binom{p^{w}}{i} \\
&\leq \frac{1}{\binom{p^{w}+y}{p^{w}}} \frac{y}{p^{w}+y-1} \sum^{y}_{i=1} \binom{y}{i} \binom{p^{w}}{i} \\
& <\frac{1}{\binom{p^{w}+y}{p^{w}}} \frac{y}{p^{w}+y-1} \sum^{y}_{i=0} \binom{y}{i} \binom{p^{w}}{i} = \frac{y}{p^{w}+y-1}.
\end{align*}
Hence, $\frac{y-2}{p^{w}+y-1} < L_{\rm Li} < \frac{y}{p^{w}+y-1}$ as $p\to +\infty$. Thus,
\[
\lim_{p\to +\infty} \frac{(y-2)p^{w}-2y} {yp^{w}+y^{2}-y} < \lim_{p\to +\infty} \frac{L_{\rm Li}}{L_{1}} < \lim_{p\to +\infty} \frac{p^{m}+y}{p^{m}+y-1}.
\]
Since, as $p\to +\infty$, $\lfloor\frac{w}{2}\rfloor \geq \max\{v_{1},v_{2},\ldots,v_{c}\}$ and $y >>2$, we have
\[
\lim_{p\to +\infty} \frac{(y-2)p^{w}-2y} {yp^{w}+y^{2}-y} = \lim_{p\to +\infty} \frac{p^{m}+y}{p^{m}+y-1} = 1 \mbox{, yielding }
\lim_{p\to +\infty}\frac{L_{\rm Li}}{L_{1}} = 1.
\]

\begin{figure*}[ht!]
\caption{Comparison of parameters between our scheme with the scheme of \cite{li2017}.}
\centering
\begin{tabular}{cc}
\includegraphics[width=0.45\linewidth]{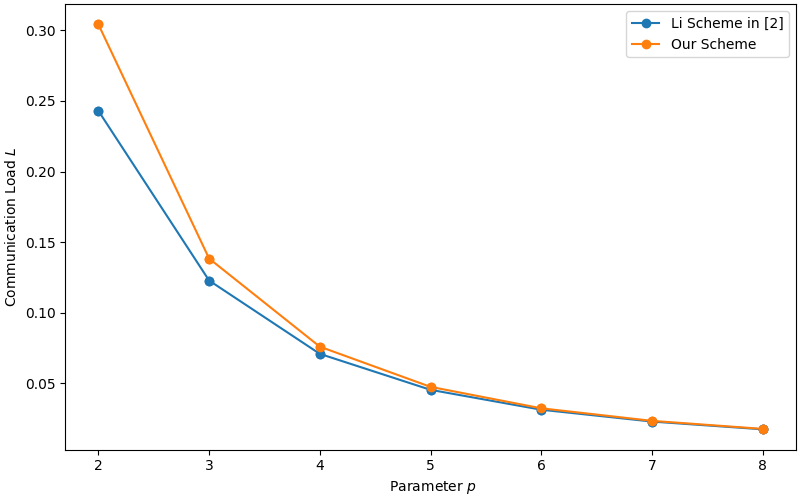} & 
\includegraphics[width=0.45\linewidth]{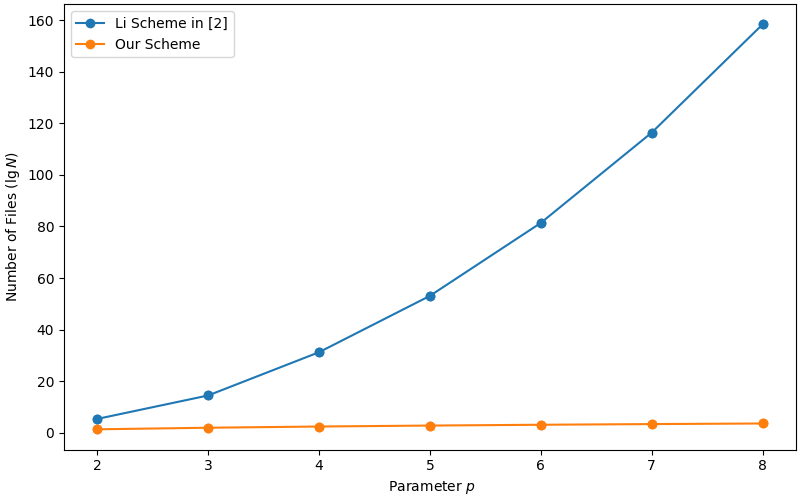}\\
(a) Communication loads & 
(b) Number of input files (in $\log$)\\
\end{tabular}
\label{fig3}
\end{figure*}

For $K = p^{4} + p^{2}+p+1$ and $r=s=p^{4}$, Figure \ref{fig3} compares our scheme with that of Li {\it et al.} in \cite{li2017} in terms of communication load and the number of input files.

\begin{remark} \label{rem1}
Fixing $w=2$ and $y=p+1$, for a given prime power $p$, we obtain the cascaded CDC scheme that has the same numbers of input files, output functions, nodes, $r$, and $s$ with those of the first scheme in \cite{jiang2022}. Our construction, therefore, obtain a new class of asymptotically optimal cascaded CDC schemes with more flexible parameters.
\end{remark}

\begin{remark} We briefly revisit the cascaded CDC schemes constructed in \cite{cheng2023} by using $t$-designs and $t$-GDDs. When the number of nodes $K$ is large and $r=s$, all of their respective communication loads approach $\frac{1}{2}$, which is clearly larger than the one in our scheme. This suffices to confirm that the scheme in Theorem \ref{thm1} performs better.
\end{remark}

\section{Concluding Remarks}\label{sec:conclude}
We have just presented a new class of cascaded coded distributed computing schemes by using $1$-designs. The parameters and construction method have been given in Theorem \ref{thm1} and its proof. 

When compared with the first scheme in \cite{jiang2022}, ours has less communication load, all other parameters being equal. Figure \ref{fig2}) illustrates this fact. 

In comparison with  the scheme in \cite{li2017}, ours has less numbers of input files and output functions when the variables $r$ and $s$ are kept the same. Asymptotically, when the number of nodes $K$ is sufficiently large, the communication load of our scheme approaches that of the CDC scheme in \cite{li2017}. The trend can be clearly seen, already for relatively small values of the parameter $p$, in Figure \ref{fig3}. We have also showed that our new class of asymptotically optimal cascaded CDC schemes has more flexible parameters than the first scheme in \cite{jiang2022}. Remark \ref{rem1} summarizes this observation.

\section*{Appendix A}\label{AppendixA:PfThm2}
\centerline{Proof of Theorem \ref{thm1}}
We prove Theorem \ref{thm1} by explicitly constructing the claimed cascaded CDC scheme. 

Let $\mathcal{X}=[0,n-1]$ and let $\mathcal{B}_{0}=\{0,1,\ldots,t-1\}$, with $3t\geq 2n$ and $n>t$. For any $i \in [n-1]$, let $\mathcal{B}_{i}=\{c_{1},c_{2},\ldots,c_{t}\}$, where $c_{j} \equiv (i-1+j) \pmod{n}$, with $j \in [t]$. Denoting $\mathfrak{B}=\{\mathcal{B}_{0},\mathcal{B}_{1},\ldots,\mathcal{B}_{n-1}\}$, we confirm that $(\mathcal{X},\mathfrak{B})$ is a $1$-design with parameters $(n,t,t)$, whereas $(\mathcal{X},\mathfrak{B})$ is \emph{not} a $2$-design.

\begin{example}
Let $\mathcal{X}=[0,4]$ and let $\mathfrak{B}=\{\{0,1,2\},\{1,2,3\},\{2,3,4\},\{3,4,0\},\{4,0,2\}\}$. We verify that $(\mathcal{X},\mathfrak{B})$ is a $1$-design with parameters $(5,3,3)$. It is not a $2$-design since the pair $\{0,1\}$ is contained in a single element of $\mathfrak{B}$, but the pair $\{0,2\}$ is contained in two elements of  $\mathfrak{B}$.
\end{example}

We proceed to construct a CDC scheme with $n$ nodes, $\mathcal{K}=\mathfrak{B}$, on the $n$ files in $W=\{w_{0},w_{1},\ldots,w_{n-1}\}$ and the $n$ functions in $\mathcal{Q}=\{\phi_{0},\phi_{1},\ldots,\phi_{n-1}\}$.

In the {\tt Map} phase, let each node $\mathcal{B}_{c}\in \mathfrak{B}$, with $ c \in [0,n-1]$, store files in $Z_{\mathcal{B}_{c}} = \{w_{x} \, : \, x\in\mathcal{B}_{c}\}$. Since the cardinality of any block is $\vert \mathcal{B}_{c}\vert = t$, the computation load is
\[
r=\frac{\sum^{n-1}_{c=0}\vert Z_{\mathcal{B}_{c}}\vert}{n}=t.
\]
Using the stored files and the map functions, node $\mathcal{B}_{c}$ can compute the intermediate values in
\[
\mathcal{I}_{\mathcal{B}_{c}}=\{v_{y,x}=g_{y,x}(w_{x}) \, : \, y\in \mathcal{X} \mbox{ and } x\in \mathcal{B}_{c}\}.
\]

In the {\tt Shuffle} phase, let each node $\mathcal{B}_{c}\in \mathfrak{B}$ be arranged to compute the {\tt Reduce} functions in
\[
\mathcal{Q}_{\mathcal{B}_{c}}=\{u_{y}=\phi_{y}(w_{0},w_{1},\ldots,w_{n-1}) \, :  y\in \mathcal{B}_{c}\}.
\]
For any $x,y\in \mathcal{X}$ and any block $\mathcal{B}_{c}\in \mathfrak{B}$, the intermediate value $v_{y,x}$ is required but cannot be locally computed by $\mathcal{B}_{c}$ if and only if $y\in \mathcal{B}_{c}$ and $x\notin \mathcal{B}_{c}$. On the other hand, $v_{y,x}$ is locally computable by node $\mathcal{B}_{c}$ if and only if $x\in \mathcal{B}_{c}$. Letting $i\in \mathcal{X}$, there exist $t$ nodes that, each, stores $w_{i}$, and $n-t$ nodes that, each, does not store $w_{i}$. Without loss of generality, let $i=n-1$. In such a case, none of the nodes in $\{\mathcal{B}_{j} \, : \, j \in [0,n-t-1]\}$ stores the file $w_{n-1}$. Each node $\mathcal{B}_{j}$ has access to the $t$ stored files in $\{w_{j},w_{j+1},\ldots,w_{j+t-1}\}$. Hence, node $\mathcal{B}_{j}$ has the intermediate values in
\[
\mathcal{V}^{n-1}_{j} = \{v_{n-1,j},v_{n-1,j+1},\ldots,v_{n-1,j+t-1}\}.
\]
If $1,a_{1},a_{2},\ldots,a_{2t-1-n}\in\mathbb{F}_{2^{T}}$ are all distinct and $a_{i}\neq 0$ for any $1\leq i\leq 2t-1-n$, then $2^{T}\geq 2t-n+1$. The nodes in $\{\mathcal{B}_{j} \, : \, j \in [0,n-t-1]\}$ collectively multicast the $n-t$ signals
\begin{align*}
X_{\mathcal{B}_{0}}[1]&=v_{n-1,0}+ v_{n-1,1}+\ldots+v_{n-1,n-t-1}+ v_{n-1,n-t}+\ldots+v_{n-1,t-2}+v_{n-1,t-1}, \\
X_{\mathcal{B}_{1}}[2]&=v_{n-1,1}+\ldots+ v_{n-1,n-t-1}+a_{1}v_{n-1,n-t}+ \ldots+ a_{2t-1-n} v_{n-1,t-2}+v_{n-1,t-1}+v_{n-1,t}, \\
&~~\vdots \\
X_{\mathcal{B}_{n-t-1}}[n-t]&=v_{n-1,n-t-1}+a^{n-1-t}_{1}v_{n-1,n-t}+\ldots+a^{n-1-t}_{2t-1-n}v_{n-1,t-2}+v_{n-1,t-1}+\ldots+v_{n-1,n-2},
\end{align*}
which we can express as
\[
\begin{pmatrix}
X_{\mathcal{B}_{0}}[1] \\
X_{\mathcal{B}_{1}}[2] \\
\vdots \\
X_{\mathcal{B}_{n-t-1}}[n-t]
\end{pmatrix}
= 
\begin{pmatrix}
1 & 1 & \cdots & 1 & 1 & \cdots & 1 & 1 & 0 & \cdots & 0\\
0 & 1 & \cdots & 1 & a_{1} & \cdots & a_{2t-1-n} & 1 & 1 & \cdots & 0\\
\vdots & \vdots & \ddots & \vdots & \vdots & \ddots & \vdots & \vdots & \vdots & \ddots & \vdots\\
0 & 0 & \cdots & 1 & a^{n-1-t}_{1} & \cdots & a^{n-1-t}_{2t-1-n} & 1 & 1 & \cdots & 1
\end{pmatrix} 
\begin{pmatrix}
v_{n-1,0} \\
v_{n-1,1} \\
\vdots \\
v_{n-1,n-2}
\end{pmatrix}.
\]
The total number of bits pertaining $v_{n-1,y}$ transmitted by any node $\mathcal{B}_{j}$, for $j \in [0, n-t-1]$, is $(n-t) T$. This translates into $n-t$ intermediate values. Thus, the total number of intermediate values $v_{x,y}$ transmitted by all the nodes combined is $(n-t)n$.

If a node $\mathcal{B}_{m}$ is arranged to compute $\phi_{n-1}$ but is unable to compute $v_{n-1,x}$, then $w_{n-1}\in\mathcal{B}_{j}$ and $w_{x}\notin\mathcal{B}_{j}$. It is easy to see that $n-t \leq m\leq n-1$. We divide these nodes into five cases.

\textbf{Case 1:} If $m=n-t$, then, by the definition of $\mathcal{B}_{n-t}$, the node $\mathcal{B}_{n-t}$ can locally compute
\[
v_{n-1,n-t},v_{n-1,n-t+1},\ldots,v_{n-1,n-2}.
\]
Thus, $\mathcal{B}_{n-t}$ only needs to compute the system of equations
\[
\begin{pmatrix}
X_{\mathcal{B}_{0}}[1]-\sum^{t-1}_{i=n-t}v_{n-1,i} \\
X_{\mathcal{B}_{1}}[2]-\sum^{t-2}_{i=n-t}a_{i-n+t+1}v_{n-1,i}-v_{n-1,t-1}-v_{n-1,t} \\
\vdots \\
X_{\mathcal{B}_{n-t-1}}[n-t]-\sum^{t-2}_{i=n-t}a^{n-1-t}_{i-n+t+1}v_{n-1,i}-\sum^{n-2}_{i=t-1}v_{n-1,i}
\end{pmatrix}
=
\begin{pmatrix}
1 & 1 & \cdots & 1 \\
0 & 1 & \cdots & 1 \\
\vdots & \vdots & \ddots & \vdots \\
0 & 0 & \cdots & 1
\end{pmatrix}
\begin{pmatrix}
v_{n-1,0} \\
v_{n-1,1} \\
\vdots \\
v_{n-1,n-t-1}
\end{pmatrix}.
\]
Since the coefficient matrix is clearly nonsingular, $\mathcal{B}_{n-t}$ decodes $v_{n-1,0},v_{n-1,1}, \ldots,v_{n-1,n-t-1}$ upon receiving the signals in $\{ X_{\mathcal{B}_{i}}[i+1] \, : \, i\in [0, n-t-1]\}$.

\textbf{Case 2:} If $n-t+1 \leq m\leq 2n-2t-2$, then $\mathcal{B}_{m}$ can locally compute
\[
v_{n-1,m},\ldots,v_{n-1,n-2},v_{n-1,0},\ldots, v_{n-1,m-n+t-1}.
\]
Thus, $\mathcal{B}_{m}$ only needs to solve the system of equations
\begin{multline}\label{eq:syseq1}
\begin{pmatrix}
X_{\mathcal{B}_{0}}[1]-\sum^{m-n+t-1}_{i=0}v_{n-1,i}-\sum^{t-1}_{i=m}v_{n-1,i} \\
X_{\mathcal{B}_{1}}[2]-\sum^{t-2}_{i=m}a_{i-n+t+1}v_{n-1,i}-v_{n-1,t-1}-v_{n-1,t}-\sum^{m-n+t-1}_{i=1}v_{n-1,i} \\
\vdots \\
X_{\mathcal{B}_{n-t-1}}[n-t]-\sum^{t-2}_{i=m}a^{n-1-t}_{i-n+t+1}v_{n-1,i}-\sum^{n-2}_{i=t-1}v_{n-1,i}
\end{pmatrix} =\\
\begin{pmatrix}
1 & 1 & \cdots & 1 & 1 & \cdots & 1\\
\vdots & \vdots & \ddots & \vdots & \vdots & \ddots & \vdots \\
1 & 1 & \cdots & 1 & a^{m-n+t}_{1} & \cdots & a^{m-n+t}_{m-n+t}\\
0 & 1 & \cdots & 1 & a^{m-n+t+1}_{1} & \cdots & a^{m-n+t+1}_{m-n+t}\\
\vdots & \vdots & \ddots & \vdots & \vdots & \ddots & \vdots \\
0 & 0 & \cdots & 1 & a^{n-1-t}_{1} & \cdots & a^{n-1-t}_{m-n+t}
\end{pmatrix}
\begin{pmatrix}
v_{n-1,m-n+t} \\
v_{n-1,m-n+t+1} \\
\vdots \\
v_{n-1,m-1}
\end{pmatrix}.
\end{multline}
We consider the determinant 
\[
A=\begin{vmatrix}
1 & 1 & \cdots & 1 & 1 & \cdots & 1\\
\vdots & \vdots & \ddots & \vdots & \vdots & \ddots & \vdots \\
1 & 1 & \cdots & 1 & a^{m-n+t}_{1} & \cdots & a^{m-n+t}_{m-n+t}\\
0 & 1 & \cdots & 1 & a^{m-n+t+1}_{1} & \cdots & a^{m-n+t+1}_{m-n+t}\\
\vdots & \vdots & \ddots & \vdots & \vdots & \ddots & \vdots \\
0 & 0 & \cdots & 1 & a^{n-1-t}_{1} & \cdots & a^{n-1-t}_{m-n+t}
\end{vmatrix}.
\]
Node $\mathcal{B}_{m}$ can decode $v_{n-1,m-n+t}, v_{n-1,m-n+t+1},\ldots,v_{n-1,m-1}$ after receiving the signals in $\{X_{\mathcal{B}_{i}}[i+1] \, : \, i \in [0,n-t-1]\}$ if and only if $A \neq 0$. By using elementary operations, since $1,a_{1},\ldots,a_{m-n+t}$ are all distinct, we have 
\[
A =
\begin{vmatrix}
1 & 0 & \cdots & 0 & 1 & \cdots & 1\\
\vdots & \vdots & \ddots & \vdots & \vdots & \ddots & \vdots \\
1 & 0 & \cdots & 0 & a^{m-n+t}_{1} & \cdots & a^{m-n+t}_{m-n+t}\\
0 & 1 & \cdots & 0 & a^{m-n+t+1}_{1} & \cdots & a^{m-n+t+1}_{m-n+t}\\
\vdots & \vdots & \ddots & \vdots & \vdots & \ddots & \vdots \\
0 & 0 & \cdots & 1 & a^{n-1-t}_{1} & \cdots & a^{n-1-t}_{m-n+t}
\end{vmatrix}
=\begin{vmatrix}
1 &  1 & \cdots & 1\\
1 &  a_{1} & \cdots & a_{m-n+t}\\
\vdots & \vdots & \ddots & \vdots\\
1 & a^{m-n+t}_{1} & \cdots & a^{m-n+t}_{m-n+t}
\end{vmatrix} \neq 0.
\]

\textbf{Case 3:} If $2n-2t-1 \leq m\leq t$, then $\mathcal{B}_{n-t}$ can locally compute
\[
v_{n-1,m},\ldots,v_{n-1,n-2},v_{n-1,0},\ldots, v_{n-1,m-n+t-1}.
\]
Thus, $\mathcal{B}_{m}$ only needs to focus on the system of equations
\begin{multline}\label{eq:syseq2}
\begin{pmatrix}
X_{\mathcal{B}_{0}}[1]-\sum^{m-n+t-1}_{i=0}v_{n-1,i}-\sum^{t-1}_{i=m}v_{n-1,i} \\
X_{\mathcal{B}_{1}}[2]-\sum^{m-n+t-1}_{i=n-t-1}a_{i-n+t+1}v_{n-1,i}-\sum^{t-1}_{i=m}a_{i-n+t+1}v_{n-1,i}-v_{n-1,t}-\sum^{n-t-2}_{i=1}v_{n-1,i} \\
\vdots \\
X_{\mathcal{B}_{n-t-1}}[n-t]-\sum^{m-n+t-1}_{i=n-t-1}a^{n-1-t}_{i-n+t+1}v_{n-1,i}-\sum^{t-1}_{i=m}a^{n-1-t}_{i-n+t+1}v_{n-1,i}-\sum^{n-2}_{i=t}v_{n-1,i}
\end{pmatrix} = \\
\begin{pmatrix}
1 & 1 & \cdots & 1 \\
a_{m-2n+2t+1} & a_{m-2n+2t+2} & \cdots & a_{m-n+t} \\
\vdots & \vdots & \ddots & \vdots\\
a^{n-1-t}_{m-2n+2t+1} & a^{n-1-t}_{m-2n+2t+2} & \cdots & a^{n-1-t}_{m-n+t} \\
\end{pmatrix}
\begin{pmatrix}
v_{n-1,m-n+t} \\
v_{n-1,m-n+t+1} \\
\vdots \\
v_{n-1,m-1}
\end{pmatrix},
\end{multline}
where $a_{0}=1$ and $a_{2t-n}=1$. The coefficient matrix is clearly Vandermonde. Since $a_{m-2n+2t+1}, a_{m-2n+2t+2},\ldots,a_{m-n+t}$ are all distinct, $\mathcal{B}_{m}$ decodes $v_{n-1,m-n+t},v_{n-1,m-n+t+1},\ldots,v_{n-1,m-1}$ upon receiving the signals $X_{\mathcal{B}_{i}}[i+1]$, for $i \in [0,n-t-1]$.

\textbf{Case 4:} The case of $t+1 \leq m \leq n-2$ can be proven in a similar way as in the proof of Case 2 above.

\textbf{Case 5:} The case of $m=n-1$ can be proven in a similar way as in the proof of Case 1 above.

Now that all cases have been settled, we can conclude that the communication load is indeed 
\[
L=\frac{n \, (n-t) \, T}{Q \cdot N \cdot T} = \frac{n-t}{n}.
\]

In the {\tt Reduce} phase, building on the {\tt Shuffle} phase, each node $\mathcal{B}\in \mathfrak{B}$ can derive the intermediate values in
\[
\{v_{x,y} \, : \,  x\in\mathcal{B}, y\notin\mathcal{B}\}
\]
after receiving the signals from all other nodes. Node $\mathcal{B}$ can then locally compute the {\tt Reduce} functions in
\[
\mathcal{Q}_{\mathcal{B}}=\{u_{y}=\phi_{y}(w_{0},w_{1},\ldots,w_{n-1}) \, : y\in\mathcal{B}\}.
\]
Our proof is now complete. \qed

\section*{Appendix B}\label{appendixB:PfLem3}
\centerline {Proof of Lemma \ref{lem3.1}}

We begin by showing that 
\begin{equation} \label{for32}
2 \binom{y}{3} \binom{p^{w}}{y-3} > 
\sum^{y-3}_{i=0} (y-2-i) \binom{y}{y-i} \binom{p^{w}}{i}.
\end{equation}
For any $i\in [0,y-3]$, let
\[
b_{i} = (y-2-i) \binom{y}{y-i} \binom{p^{w}}{i}.
\]
Hence, for any $i\in [0,y-4]$,
\[
\frac{b_{i}} {b_{i+1}} = \frac{y-2-i}{y-3-i} \frac{(i+1)^{2}}{(p^{w}-i)(y-i)},
\]
which is an increasing function in the stated range of $i$, making 
\[
\frac{b_{i}}{b_{i+1}} \leq 
\frac{b_{y-4}}{b_{y-3}} = 
\frac{1}{2}\frac{(y-3)^{2}}{p^{w}-y+4}.
\]
Since $\lfloor\frac{w}{2}\rfloor \geq \max\{v_{1},v_{2},\cdots,v_{c}\}$, the highest order term of $y^{2}$ about $p$ is $\leq w$, which convinces us that 
\[
\lim_{p\rightarrow \infty} \frac{(y-3)^{2}}{p^{w}-y+4} \leq 1.
\]
Hence, 
\[
\frac{b_{i}}{b_{i+1}} \leq 
\frac{b_{y-4}}{b_{y-3}} \leq \frac{1}{2}
\]
and, for any $i \in [0,y-4]$,
\[
b_{i}\leq\frac{1}{2}b_{i+1}\leq \ldots \leq(\frac{1}{2})^{y-3-i}b_{y-3}.
\]
Thus, 
\[
\sum^{y-3}_{i=0}b_{i}  \leq 
\sum^{y-3}_{i=0} \left(\frac{1}{2}\right)^{y-3-i}b_{y-3}  =\left(2-\left(\frac{1}{2}\right)^{y-3}\right) b_{y-3}<2b_{y-3},
\]
from which, since we now have $2b_{y-3} > \sum^{y-3}_{i=0}b_{i}$, we arrive at the desired inequality in \eqref{for32}.
  
Our next task is to establish 
\begin{equation}\label{for33}
\binom{p^{w}}{y} > \binom{y}{3} \binom{p^{w}}{y-3}.
\end{equation}
We can directly prove that 
\[
\frac{\binom{p^{w}}{y}}{\binom{y}{3} \binom{p^{w}}{y-3}}= \frac{6\sum^{2}_{j=0}(p^{w}-y+1+j)}{\sum^{2}_{j=0}(y-j)^{2}}.
\]
Since $\lfloor\frac{w}{2}\rfloor \geq \max\{v_{1},v_{2},\ldots,v_{c}\}$, the highest order term of $y^{2}$ about $p$ is $\leq w$, implying 
\[
\lim_{p\to \infty} \frac{p^{w}-y+1+j} 
{(y-j)^{2}} \geq 1 \mbox{ for any } j \in \{0,1,2\}.
\]
Hence, 
\[
\frac{6\sum^{2}_{j=0}(p^{w}-y+1+j)} {\sum^{2}_{j=0}(y-j)^{2}} > 1 \mbox{, yielding } 
\binom{p^{w}}{y} > \binom{y}{3} \binom{p^{w}}{y-3}. 
\]

Now that we have done the preparatory work, the inequality in \eqref{for31} can soon be confirmed. Using \eqref{for32} and \eqref{for33}, we infer 
\begin{align*}
\sum^{y}_{i=0} i \binom{y}{y-i} \binom{p^{w}}{i}
&=\sum^{y-3}_{i=0} i \binom{y}{i} 
\binom{p^{w}}{i} + (y-2) \sum^{y}_{i=y-2} \binom{y}{y-i} \binom{p^{w}}{i} + \binom{y}{1} \binom{p^{w}}{y-1} + 2 \binom{p^{w}}{y} \\
& > \sum^{y-3}_{i=0} i \binom{y}{i} \binom{p^{w}}{i} + (y-2) \sum^{y}_{i=y-2} \binom{y}{y-i} \binom{p^{w}}{i} + 2 \binom{p^{w}}{y} \\
& > \sum^{y-3}_{i=0} i \binom{y}{i} \binom{p^{w}}{i} + (y-2) \sum^{y}_{i=y-2} \binom{y}{y-i} \binom{p^{w}}{i} + 2 \binom{y}{3} \binom{p^{w}}{y-3} \\
& > (y-2) \sum^{y-3}_{i=0} \binom{y}{y-i} \binom{p^{w}}{i} + (y-2) \sum^{y}_{i=y-2} \binom{y}{y-i} \binom{p^{w}}{i}\\
& = (y-2) \binom{p^{w}+y}{y}. \qed
\end{align*}


\end{document}